\documentclass[12pt]{article}
\usepackage[a4paper, margin = 1.2in]{geometry}
\usepackage{graphicx} 
\usepackage{amssymb}
\usepackage{xcolor}
\usepackage{amsmath}
\usepackage{amsthm}
\usepackage{bbm}
\usepackage{caption}
\usepackage{subcaption}
\usepackage{url}

\usepackage{breakurl}
\usepackage[breaklinks]{hyperref}
\usepackage[toc, page]{appendix}
\usepackage{cite}
\usepackage{authblk}

\newtheoremstyle{example}{}{}{\em}{}{\bfseries}{\smallskip}{\newline}{}
\theoremstyle{example}

\title{Realigning Incentives to Build Better Software: A Holistic Approach to Vendor Accountability}
\author[1,3]{Gergely Biczók}
\author[2]{Sasha Romanosky}
\author[3]{Mingyan Liu}
\affil[1]{CrySyS Lab, Budapest Univ. of Technology and Economics\\\texttt{biczok@crysys.hu}}
\affil[2]{RAND Corporation\\\texttt{sromanos@rand.org}}
\affil[3]{University of Michigan\\\texttt{mingyan@umich.edu}}
\date{}

\ifodd 1
\newcommand{\com}[1]{{\color{red}\textbf{Comment}: #1}}
\newcommand{\resp}[1]{{\color{cyan}\textbf{Response}: #1}} 
\else

\newcommand{\com}[1]{}
\newcommand{\resp}[1]{}
\fi

\begin{document}

\maketitle

\begin{abstract}
In this paper, we ask the question of why the quality of commercial software, in terms of security and safety, does not measure up to that of other (durable) consumer goods we have come to expect. 
We examine this question through the lens of incentives. We argue that the challenge around better quality software is due in no small part to a sequence of misaligned incentives, the most critical of which being that the harm caused by software problems is, by and large, shouldered by consumers, not developers.  This lack of liability means software vendors have every incentive to rush low-quality software onto the market and no incentive to enhance quality control. Within this context, this paper outlines a holistic technical and policy framework we believe is needed to incentivize better and more secure software development. At the heart of the incentive realignment is the concept of software liability. This framework touches on various components, including legal, technical, and financial, that are needed for software liability to work in practice; some currently exist, some will need to be re-imagined or established.  This is primarily a market-driven approach that emphasizes voluntary participation but highlights the role appropriate regulation can play. We connect and contrast this with the EU legal environment and discuss what this framework means for open-source software (OSS) development and emerging AI risks. Moreover, we present a CrowdStrike case study complete with a what-if analysis had our proposed framework been in effect. Our intention is very much to stimulate a robust conversation among both researchers and practitioners.
\end{abstract}

\section{Introduction}\label{sec:intro}

\subsection{Background and Motivation} 

It is not unreasonable to ask ``Why can't software firms make better software?'' or ``Why can't software be better regulated like other consumer goods?'' Here, by {\em better}, we loosely mean quality on par with the type of reliability, performance, and safety features we routinely expect of other (durable) consumer goods we purchase and use in our daily lives: appliances, furniture, vehicles, etc. 

We contend that there are fundamental problems with the software development and deployment ecosystem and that these problems will only grow with time unless fundamental changes are made.  Globally, we interact with more software applications across more consumer and commercial devices that operate and influence more critical parts of our lives. Over time, these software products have been expanding in size and becoming more complex, in part from referencing and incorporating more third-party libraries and dependencies. Every additional library makes it more and more difficult to identify design, logic, and implementation vulnerabilities. 
Indeed, firms may well have no idea about existing or newly discovered vulnerabilities in the third-party code they use and deploy.
For example, the widely discussed 2021 log4j incident left companies scrambling to assess whether their own systems are vulnerable owing to the popularity and large-scale reuse of the log4j package via the software supply chain~\cite{log4j_measure}. The ineffectiveness of the lengthy assessment enabled attackers to continue exploiting the Log4Shell vulnerability throughout the next year. With nearly $30,000$ vulnerabilities disclosed in 2024, and increasing monotonically year over year, such incidents are bound to happen even more frequently~\cite{cve_website_2025}. 

Software problems, of course, go well beyond vulnerabilities. Users of modern commercial networked software applications also very often lack awareness of the relative security and safety of these software products and how exposed they may or may not be to newly discovered vulnerabilities. 
Overall, the cost of ``poor quality'' software (a term encompassing a wide range of non-functional attributes such as security) was reported by the Consortium for Information \& Software Quality (CISQ) to be in the trillions in 2020~\cite{cisq_report}.

Many scholars have argued that the current market does not enforce sufficient accountability on the software or hardware vendor~\cite{lawfare_Dempsey24, lawfare_Dempsey24b, lawfare_Lipner24, lawfare_Bambauer24}. Indeed, many residential and commercial devices are sold without appreciation for security and privacy concerns, or often without any expectation of providing fixes for their software as new vulnerabilities are disclosed.
At the same time, 
the current U.S. legal regime is not capable of holding vendors liable for any losses from faulty or improperly maintained software, economic or otherwise. Consequently, harms are passed downward to end-users and consumers. 

Different efforts have attempted to mitigate challenges with faulty software. For example, vulnerability disclosure~\cite{arora2008optimal} and bug bounty programs~\cite{10.1093/cybsec/tyx008} have emerged as effective tools for crowdsourcing vulnerability hunting to security researchers, enabling subsequent patch development by software vendors and, finally, patching by system administrators.
As successful as these efforts have been, they only address the symptoms rather than provide a cure to the problem. In addition, NIST, BSA, Microsoft, and others have developed secure coding standards (e.g.,~\cite{bsa_2025, msdf_SDL_2025, safecode_2024, ssdf_2021}), which could presumably prevent many of the problems, to begin with; however, they are not enforced and therefore are not systematically employed. Worse, firms often deliberately eschew such best practices in order to cut costs and minimize time to market.

Collectively, we argue that this expanding software universe, combined with difficulties in managing the growing software codebase, the lack of accountability and liability, and partial solutions that address symptoms rather than root causes, together create more opportunities for both accidental failures and malicious attacks, increasing the risk caused by these software components. The importance and non-triviality of this issue and its policy context have been recognized by the broader research community, as illustrated by its inclusion in the top 10 ``cyber hard problems'' compiled by the U.S. National Academy of Sciences~\cite{national2025cyber} (Problem 5).


\subsection{Misaligned incentives and proposed realignment}
\label{sec:incentives}

And so, how can we produce ``better'' software? In this paper, we examine this question through the lens of {\em incentives}.  We argue that the challenge around better software is due in no small part to a sequence of misaligned incentives. Reliability, safety, and security features in a software product can appear invisible to the user and, thus, hard for the software developer to monetize. Therefore, spending resources on improving these features may never lead to immediate or substantial returns. Moreover, the first-mover advantage incentivizes developers and vendors to release (new versions of) products quickly without fixing (sometimes known) vulnerabilities. 

Perhaps most importantly, the cost of these vulnerabilities is largely borne by the buyers and users of the software. While there are costs associated with developing patches, vendors do not bear the cost of these (in)actions; it is very much left to the users to absorb the cost of deploying the patch (or upgrading the software), which can be substantial when it entails testing and system downtime. 
Furthermore, any harm stemming from software is typically waived by the user as part of the purchase agreement\footnote{see, e.g.,  ``The overall consensus among legal experts is that CrowdStrike is likely protected by its terms and conditions (\url{https://www.crowdstrike.com/terms-conditions/}) from reimbursing customers for more than they paid for the product.''~\cite{crowd2024}}; thus, the vendor has no incentive to minimize vulnerabilities prior to delivering the software.  As a result, the ``damage'' a software vendor is responsible for is typically, at most, the cost of the software or software contract rather than the potential harm a faulty software product can inflict. This has contributed to the lax approach to risk mitigation owing to moral hazard~\cite{DBLP:conf/acsac/Anderson01}. 

\begin{figure}[tb]
\vspace{-10pt}
\centering
\includegraphics[width=0.99\textwidth]{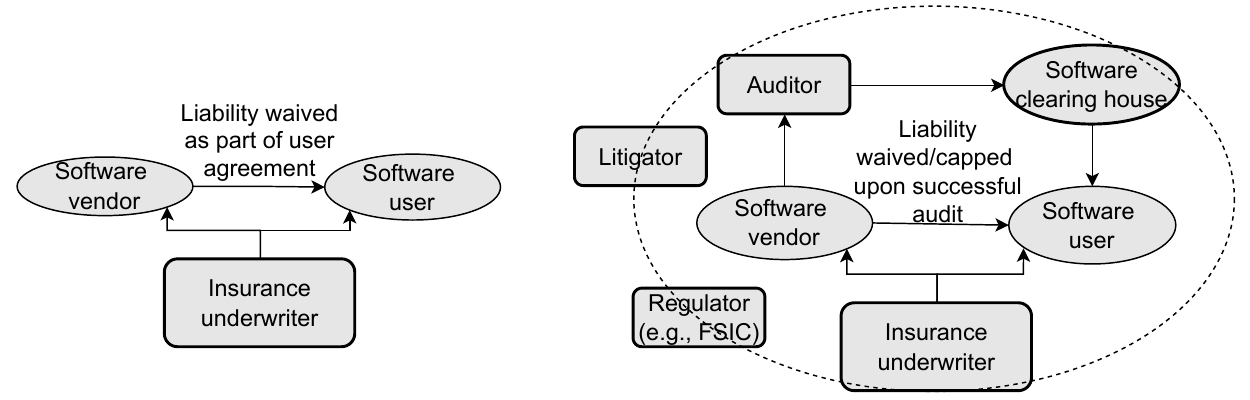}
\caption{The current (L) and proposed (R) software development and consumption ecosystem.}
\label{fig:overview-new}
\end{figure}

In short, software developers/vendors have every incentive to rush low-quality software onto the market, which is misaligned with the consumer's incentive to have secure/high-quality software for their money.
Against this backdrop, this paper outlines a technical and policy framework for an ecosystem we believe is needed to incentivize better and more secure software development. At the heart of the incentive realignment is the concept of software liability. Together with greater transparency, we believe it can realign the vendor's incentives in the direction favoring the consumers -- and potentially welcomed by society.
We shall primarily focus on software vendors: companies that develop, test, and deploy software within commercial transactions. Vendors have the greatest control over software security and safety, are the least cost avoider~\cite{gilles1992negligence}, and are uniquely positioned to serve as key gatekeepers in building a more resilient software ecosystem. In Figure~\ref{fig:overview-new}, we contrast the relatively simple current system of software supply and demand (left panel) with the ecosystem we envision (right panel). We advocate for various changes/augmentations to each of the latter's components aimed at aligning incentives for better software development. 

While many of the elements in the proposed ecosystem currently exist, others may need to be established or re-imagined.  Some address ex-ante issues (what happens prior to the software development or adoption), and some ex-post (what happens when liability issues emerge following the revelation of software vulnerabilities to which business loss can be attributed). Above all, we argue that the incentives among these components need to be (re)aligned for the ecosystem to function properly. Each component can provide value on its own, but we argue that collectively, they can help create an effective system with the right incentives and checks and balances to induce more vendor responsibility in producing secure software. We discuss each component and the incentive relationships in detail in Section~\ref{sec:system}. 

We emphasize ``incentives'' because our ultimate goal is to conceptualize a system where incentives are meaningfully aligned for market mechanisms to work effectively, where participation in any and all of these components by a firm (either vendor or customer) is entirely optional, and where regulatory tools are necessary but not the dominant component. Such a system may be better suited for the United States than for the European Union, where the regulatory power is much more prominent, as seen in the implementation of GDPR and in the most recent slew of software liability regulations (which we review briefly in Section~\ref{sec:state} and in detail in Appendix~\ref{sec:literature}). For the same reason, we do not necessarily advocate for mandating any of the components in this system; rather, our hope is that with the right incentives, the market can steer stakeholders to participate.

Importantly, we do not assume that the better alignment of incentives will eradicate all software flaws and software-based cyber incidents. Nor do we advocate for heavy-handed government regulation to oversee and interfere with commercial software development and deployment. Instead, we seek to introduce mechanisms in order to bolster accountability, reduce externalities, and balance information asymmetries within the software development and deployment ecosystem. 

The discussion presented in this paper primarily serves as an opening gambit at this stage. Our intention is very much to stimulate a robust conversation among both researchers and practitioners. 

\subsection{Organization of the paper}
\label{sec:org}
In the remainder of the paper, we first provide a streamlined overview of the current regulatory landscape around software in Section~\ref{sec:state}. In Section~\ref{sec:system}, we dive into the proposed framework and elaborate on each component and its role in the larger ecosystem. The recent CrowdStrike fiasco is presented in Section~\ref{sec:casestudies} as a case study to highlight how things might have worked differently had such an ecosystem been in place. We discuss the issue of open-source software (OSS) and AI software liability in the context of this framework in Section ~\ref{sec:discussion}. We provide a detailed literature review around recent regulatory activity and debate in software liability and transparency, both in the EU and in the US, in Appendix~\ref{sec:literature}. We conclude the paper in Section \ref{sec:conclusion}. 


%
%

\section{Current State of Software Regulation}
\label{sec:state}
Below, we review the developments in the regulatory landscape most relevant to the present paper. A more comprehensive review is provided in Appendix~\ref{sec:literature} for the interested reader.

The most obvious tool used to regulate software is liability.
Liability as a legal concept aims at establishing legal obligations for individuals and organizations to assume responsibility for their actions, particularly when such actions result in harm or damage to others. Liability encompasses various legal principles and frameworks that determine when (duty of care) and to what extent (standard of care) one party may be held accountable for the consequences of their behavior. 

Historically, software companies frequently avoided product liability using a combination of legal gray zones and disclaimers, capitalizing on the broad interpretation of acceptable user risk. As mentioned earlier, the contractual obligation of a software vendor to the software buyer/user is typically capped at the value of the contract (the cost to purchase the software) and not the potential harm caused by said software.  The typical insurance policy of the software vendor accordingly covers only this very limited liability (while separate cyber policies will cover data breaches and other cyber losses more broadly). 

Legal and regulatory developments in this space can be categorized in terms of whether software should be treated as a product, a process, or both, whether the \textit{quality} of software should be assessed by its end product or its production process, or both, whether this assessment should be mandatory or voluntary, and whether mere self-disclosure (more transparency into the product and the process) is adequate.  We briefly touch on each of these below. 

\paragraph{Software as a product. } 
The EU Council is arguably at the forefront of the ``product'' movement: it enacted a new Product Liability Directive (PLD)~\cite{PLD2024} that came into force in December 2024, with the intent to explicitly designate software (including AI software) a ``product''; this in turn comes with strict liability clauses. The PLD allows individuals, including consumers, to seek compensation from manufacturers on a strict liability basis for defective products and, in some cases, their components in the EU market. Its goal is to simplify the claims process for damages caused by product defects. It expands liability to nearly all supply chain operators and ensures consumer protection regardless of a product's origin (EU or non-EU). Consumers can seek compensation in complex cases, including safety/security regulation breaches (the draft AI Liability Directive (AILD~\cite{AILD}) further specifies such motions in the context of AI software\footnote{AILD draft revoked:~\url{https://www.euractiv.com/section/tech/news/commission-plans-to-withdraw-ai-liability-directive-draw-mixed-reactions/}}). The removal of arbitrary thresholds allows full compensation for damages. With a very broad definition of product, this PLD addresses new technologies such as cyber vulnerabilities, essential digital services, and software updates. Software components include any item (e.g., a software library), raw material, or service (e.g., when software initiates remote calls to a SaaS instance) integrated into or connected with a product, though standalone services are generally not covered. Notably, free and open-source software (OSS) outside commercial activity is excluded. 

\paragraph{Software as (mostly) a service. } 
Software is not classified as a product or product component under current US law \cite{lubin_sw_2024}. This has to do with the perceived intangible nature of software vs. the definition of product as ``tangible personal property distributed commercially for use or consumption.'' Court rulings largely align with this view by treating software as a service rather than a product, forcing plaintiffs to rely on legal arguments of negligence liability when bringing private actions against software providers. Duties and standards of care are hard to establish since the same software code could be used in different applications and use cases.  This legal backdrop notwithstanding, in several recent cases, software (e.g., the Lyft app) or components of software (e.g., parental control functionality in social media apps) were deemed as products/component products based on explanations that the definition of ``product'' should accommodate technology developments and that intangible personal property might even qualify as such. These decisions might signal a judicial change. 

\paragraph{Transparency and self-attestation. } 
While the U.S. legal community has not yet converged on the designation for software, and as a consequence, whether a liability regime is the ``right path'' forward~\cite{darkreading_Williams23}, there seems to be a consensus among scholars that attestation (of the quality of software, however it is designated) will indeed play an important role~\cite{lawfare_Dempsey24b}. Scholars argue that although market failures necessitate some form of intervention, a transparency-based approach might have the same positive effect (as that of more heavy-handed regulation) while avoiding negative consequences on cost, innovation, and competition. Mandating full disclosure of security practices removes information asymmetry and could enable consumers to assess risks and create a market-driven demand for more secure software. In theory, this would make secure development efforts financially viable (or even desired). Indeed, Lipner argues that Executive Order 14028~\cite{whitehouse_2021} has already laid out a transparency framework (based on self-attestation to the adoption of secure development practices)~\cite{lawfare_Lipner24}. Some think that attestation to a government-mandated requirement as part of procurement could incentivize vendors to improve their software development and deployment practices.

\paragraph{Safe harbor, floor, and ceiling. } 

In addition to self-attestation, a cybersecurity ``safe harbor'', based on the certification of organizations (with a similar approach already codified in the EU NIS2 Directive~\cite{NIS2}), has received considerable attention~\cite{lawfare_Tschider24}. Proponents argue that a certification safe harbor can (over time) create incentives to improve collective security practices, including supply chain security, by involving certified upstream technology providers. Dempsey further suggests a hybrid approach, combining a product-based ``floor'' for liability and a process-based ``ceiling'' for safe harbor~\cite{lawfare_Dempsey24}. His proposed mechanism for liability is three-fold. First, a rule-based approach would define the minimum standard of care for software, focusing on product features and behaviors to be included or avoided (e.g., any weaknesses tracked in the MITRE Common Weakness Enumeration (CWE) database\footnote{\url{https://cwe.mitre.org/}}). Second, as software can be complex and dynamic, a liability regime and analysis are needed to assess non-trivial design and implementation flaws; essentially, a software audit. This approach shares similarities with European legislation that has already codified voluntary~\cite{CSA2019} and mandatory~\cite{CRA2024} software audit and certification. Third, to provide a safe harbor to developers who followed secure coding recommendations, the proposal suggests a liability waiver or limitation, which makes the outcome predictable for vendors. Clearly defining the conditions under which a vendor qualifies for safe harbor provides legal certainty, reducing the need for costly and prolonged expert debates over what qualifies as reasonable (non-negligent) conduct~\cite{lawfare_Bambauer24}. A novel development in the AI arena is California Senate Bill 813~\cite{sb813}, proposing a type of safe harbor (affirmative defense) for AI developers against certain civil liability claims when systems are certified by approved and independent Multistakeholder Regulatory Organizations (MROs).
   
\section{Designing A Better Software Ecosystem }
\label{sec:system}

%
%
%
%

\begin{figure}[tb]
\centering
\includegraphics[width=0.99\textwidth]{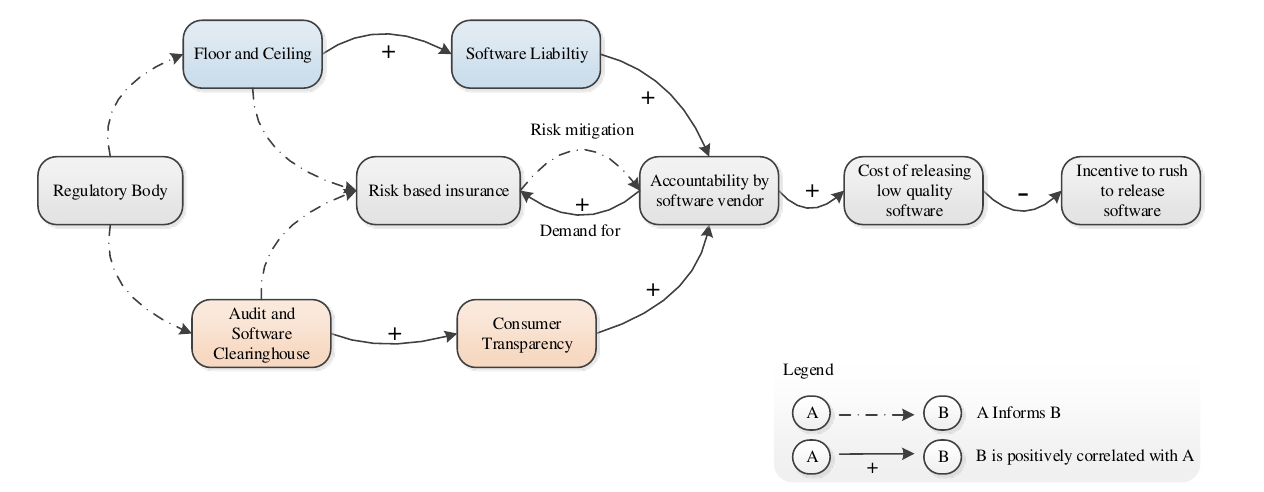}
\caption{Policies driving a better software ecosystem }
\label{fig:incentives}
\end{figure}

In this section, we describe in more detail the key components of the software ecosystem illustrated in Figure~\ref{fig:overview-new}.  As mentioned in Section~\ref{sec:intro}, we believe a stronger notion of liability for software products, supported by developments in legal, auditing, and certification services, as well as better risk-informed insurance policies, can trigger a realignment of incentives conducive to building better software. This is highlighted in Figure~\ref{fig:incentives}. 
Our proposal begins in the upper path, establishing a minimum set of software development requirements (the floor) and safe harbor practices (the ceiling). As these policies are adopted, they \textit{increase} the liability of commercial software developers for their software products. In addition, a set of process and product audits, combined with other measures, \textit{increase} transparency by signaling software quality to stakeholders. Together, these forces -- liability and transparency -- combine to \textit{increase} the accountability of vendors. This, in turn, \textit{increases} the cost of releasing low-quality (i.e., vulnerable, untested, unmaintained) software, which in turn \textit{reduces} the vendor's incentive to rush to release software. 

In addition, we believe that a software insurance mechanism can play an important role, both in aligning accountability and risk mitigation. Information about vendor coding standards and safe harbor practices, along with information about audits, can be used by insurance carriers to establish fair (i.e., risk-based) pricing. Further, as carriers become more accountable for their products, they may, at times, face costly legal actions. As their cost of software development increases, the demand for insurance policies to manage these costs will \textit{increase}.  


\subsection{Audit and Software Clearinghouse}
\label{sec:audit}


It may be argued that unless a high-quality audit system can be established in practice, this enhanced ecosystem cannot survive.  Software audit is indeed a vibrant industry with many private and for-profit consulting firms as well as independent consultants and experts engaged in this type of service.  The gaps include a) the existence of a standardized audit process, b) an accepted auditing authority, and c) an incentive for vendors to voluntarily subject themselves to an audit.  

In creating an audit standard, it's important to distinguish two parallel processes, both essential: {\em process audit} and {\em product audit}.  The former refers to checking whether the vendor has an established and documented process for software development and production, whether this process is acceptable per industry standards and best practices, whether the vendor has followed its own process, and so on.  The latter refers to checking whether the developed software contains vulnerabilities by using standard software code analysis techniques and tools; these tests are meant to catch both new vulnerabilities as well as those known to exist in previously developed software that the new software depends on. 

The audit service should, at a minimum, include the following components. (1) Completing an audit should, in general, be optional/voluntary, but may be mandatory in some specific and limited scope; e.g., certain types of software, such as those used in critical infrastructure, may be required to go through an audit. 
(2) Passing an audit waives the vendor of future liability up to some upper limit for fines or penalties it may face if/when vulnerabilities are discovered or if harm is attributed to its software. 
(3) The audit process itself should go through regular updates and improvements. Thus, the audit also carries a version number that is included in the certification it issues. 
(4) Software that fails an audit should not be allowed to be released. 

One of the key outcomes of the audit system will be a {\em Software Clearinghouse}, a repository of software that has been certified by the auditor, including libraries that feed into a myriad of other software products and systems.  In principle, software in the clearinghouse has been certified, and its subsequent use (e.g., as dependencies, clearly articulated in the software's Bill of Materials (SBOM)~\cite{zahan2023sbom}) carries a limited liability waiver as described in the previous section. We believe this will encourage customers to maximally use certified software and vendors to build on already certified dependencies. Such an approach has the potential to (partially) prevent the challenges and well-documented struggles of customers trying to identify vulnerable dependencies within their own systems upon the disclosure of newly identified vulnerabilities (especially affecting popular libraries/packages), just as it happened in the Log4Shell case~\cite{log4j_measure}.

Software product and process audits have their own set of challenges; some even question whether cybersecurity-related audits could be effective~\cite{SLAPNICAR2022100548}. In reality, software quality assurance in general and audit, in particular, is its own field of study~\cite{galin2004software} with its own standards and recommendations, e.g., IEEE 1028-2008~\cite{ieee20081028}. In relation to product-based security testing, defining a set of minimum requirements (i.e., a ``floor'') can enhance both the effectiveness of the audit and the interpretation of the resulting certification (see Section~\ref{sec:liability} for details). A ``simple'' audit similar to enforcing ``building codes'' in the construction industry could be an adequate starting point~\cite{10.1145/2700341}.


As for the technical security testing methodology, the widespread DevOps philosophy and the resulting Continuous Integration/Continuous Development (CI/CD) approach, designed to streamline and speed up software production and feature delivery, pose unique challenges. One such challenge is the need for rapid and continuous security testing, manifesting in the DevSecOps trend. Such speed-up requires automation, which is far from trivial for more powerful testing methods, e.g., dynamic testing~\cite{cicd_testing}. In the context of the proposed Software Clearinghouse, one critical enabling technology will be quick and automated (regression) testing, to keep up with the rapid releases of new versions of already certified software packages without re-doing a full audit~\cite{elbaum2014techniques}.


There are some additional steps that can be taken to ensure and enhance the transparency of software practices, more generally, in parallel to the audit process. We support the adoption of SBOMs as a way to ensure transparency concerning the key software artifacts that compose a given software application. (Note that the EU CRA already requires SBOMs in the frame of its mandatory product audit mechanism~\cite{CRA2024}.) Properly maintained SBOMs will also facilitate the audit of new software that includes components found in the clearinghouse by identifying any third-party components already certified.

Another laudable effort is put forward in the form of self-attestation in the context of the software development process~\cite{whitehouse_2021, lawfare_Dempsey24b}. In fact, there is a growing consensus among scholars and industry experts that the transparency it brings, even without a third-party independent audit, will move the needle towards adopting secure development and deployment processes and providing clarity (and a chance for differentiation) for purchasing decisions, partially alleviating the asymmetry present in current software markets. 

\subsection{Software Liability}
\label{sec:liability}

Clearly defined software audit processes can help define software liability, which can, in turn, help hold software vendors liable for low-quality software.  The two-pronged (process and product) hybrid audit process described above is closely tied to and may be implemented in the form of ``safe harbor'' vs. ``floor'' cited earlier~\cite{lawfare_Dempsey24}. The idea is that if a vendor passes a process audit, then it may be afforded a waiver for having made a good-faith effort (safe harbor). On the other hand, software that fails to meet a minimum set of product security requirements (through the product audit) fails the product audit and, thus, the overall audit, regardless of its process.  In other words, there is an agreed ``floor'' on the quality, below which no software can be certified; beyond this, as long as the process is sound, the vendor is afforded a certain level of protection.  

This approach can directly lead to a meaningful interpretation of liability, as it delineates the responsibility: the vendor is (largely) not responsible for the residual risk its product carries as long as it adheres to a specified standard of process; it is, however, wholly responsible for meeting a set of minimum quality requirements even if it has adhered to the process standards. From the user's point of view, this delineation helps them understand where their own responsibilities and risks lie: if a software has passed both the product audit and the process audit, then the user is aware that the vendor can no longer be faulted for harm that may be caused by the software. If, on the other hand, the software passed the product audit but failed the process audit, then the vendor may be responsible for compensating the user for a large share of the harm attributable to the software. In either case, the user is better informed and can use that knowledge to seek the right type of risk protection (see Section~\ref{sec:insurance} on insurance). In the same vein, this approach can help resolve legal disputes by establishing who is responsible when harm occurs.

As to how to define the minimum quality requirements, i.e., the floor, one may turn to ``unforgivable vulnerabilities''~\cite{christey2007unforgivable}. Such weaknesses are trivial to find and occur time and time again, demonstrating a ``systematic disregard'' for secure development practices. In a recent proposal, the UK National Cyber Security Centre (NCSC) proposed a method to classify vulnerabilities into ``forgivable'' and ``unforgivable'' ~\cite{ncsc_unforgivable}. The method focuses on finding the root cause of vulnerabilities and their respective top-level mitigation based on the MITRE CWE catalog. The mitigation is then assigned an ``ease of implementation'' score (taking cost, familiarity, and technical feasibility into consideration); vulnerabilities with an ``easy'' mitigation are then declared unforgivable. Simply put, such weaknesses should not appear in reasonably secure software. 

In turn, the process-based ceiling may adopt ideas from the secure software development lifecycle, e.g., Microsoft's SDL~\cite{1377211}. SDL promotes a DevSecOps approach, integrating security into DevOps (DevSecOps). It applies to all software development models and supports diverse software types (firmware, AI, IoT, web services, etc.) across multiple platforms (cloud, on-premises, mobile, SaaS, etc.). The SDL emphasizes 10 key security practices to enhance software security throughout the whole lifecycle.
Adding to this, the 2024 White House report on secure and measurable software~\cite{house2024back} further promotes the use of advanced secure technologies such as memory-safe programming languages (e.g., Rust, Go, etc.), memory-safe hardware, and formal methods to prove the correctness of the developed code. Furthermore, the report advocates for the use of cybersecurity quality metrics. Fascinatingly, it also acknowledges the need for shifting market forces, i.e., realigning incentives, to improve these metrics, which is exactly our objective in this paper.

Interestingly, the floor-ceiling approach can also help settle the two key pending legal issues concerning software liability mentioned earlier: (1) whether software will remain a ``service'' in the eyes of the law or will start being viewed as a ``product'' at some point; and (2) whether legal protections against lawsuits should be based on satisfactory process, product quality, or both. 
As far as we know, there have been no direct liability lawsuits until very recently (see Delta Airlines vs. CrowdStrike in our case study in Section~\ref{sec:casestudies}). 
This is directly tied to the fact that there is a lack of more precise definitions of liability, which is much needed.  As a reference, the EU GDPR was able to define both duty and standard of care, leading to substantial fines for mishandling personal data and prompting a reassessment of cybersecurity investments (as risk-matching cybersecurity countermeasures have also been prescribed for safeguarding personal data).  This gives one hope that similar definitions may emerge for the software industry. 

We also believe that impaneling legal professionals and litigators to examine software service level agreements (SLAs) and offer suggestions on the legality of these SLAs and whether there are ways to regulate them or, alternatively, pathways to bringing lawsuits against these SLAs, would also help with the eventual convergence.

\subsection{Software Liability Insurance}
\label{sec:insurance}
We envision that liability insurance for software vendors will play an important role in this new ecosystem. As is always the case, the primary function of insurance is to allow firms to better manage their cost of operations by transferring some of the risks to an insurer~\cite{woods2023history}. As their costs increase, particularly costs related to liability, firms will turn to carriers requesting coverage for third-party liability costs and claims. Indeed, just as with data breaches, ransomware, and the subsequent liability costs~\cite{10.7551/mitpress/13665.001.0001}, this market has emerged to adapt organically to fulfill the demand~\cite{glaring-gap-tort}.

Indeed, liability claims against software vendors are already covered by Technology Errors and Omissions (Tech E\&O) insurance policies, which cover a policyholder's third-party liability costs -- but not first-party losses incurred directly by the policyholder -- and are often combined with privacy, media, and cyber-insurance policies. These policies collectively cover losses due to typical data breaches (and associated costs), ransomware, privacy fees and violations, and, in particular, third-party liability claims due to tech products and services. 

In these policies software is typically covered as a \textbf{technology product} defined as ``created, designed, distributed, manufactured, or sold by or on behalf and for the benefit of an Insured; or Leased or licensed by an Insured to third parties,'' or a \textbf{technology service} defined as ``any computer, cloud computing, information technology, telecommunication, electronic services and any related consulting and staffing services, including data processing, data and application hosting, the provision of managed services, software as a service (SaaS), platform as a service (PaaS), infrastructure as a service (IaaS), network as a service (NaaS), computer systems analysis, computer consulting and training, programming, computer systems installation, management, repair, and maintenance, network design and Internet service''\footnote{See CyberRiskConnect, Privacy, Security and Technology Insurance, TRD 050 0619,  2019 X.L. America, Inc., though similar language exists in policies by Beazley, Willis Towers Watson, and At-Bay. Policies are available upon request.}

The policies cover costs resulting from any ``act, error, omission, neglect, negligent misrepresentation or breach of duty'' or ``failure of technology products to perform the intended function or serve their intended purpose'' or ``failure of technology services or technology products to meet any applicable legal or industry standard concerning quality, safety or fitness for a particular purpose.''\footnote{Ibid.} 

The policy exclusions are those typical of cyber-insurance policies, such as losses arising from willful or criminal acts, deceptive business practices, acts of war, patent or trade infringement, infrastructure failure (i.e., water or telecommunication), etc. In addition, these policies typically  exclude costs incurred by policyholders to ``withdraw or recall technology products, including products that incorporate an Insured’s technology products, technology services, or professional services'' or costs ``to correct, re-perform or complete any technology services or professional services.''\footnote{Ibid.} 

In terms of limits, Cowbell offers Tech E\&O Excess coverage up to \$5M in limits for mid-sized companies with revenue up to \$1B,\footnote{see \url{https://cowbell.insure/news-events/pr/cowbell-us-upmarket}}, and At-Bay provides coverage with aggregate limits of \$10M, for businesses with revenue up to \$5B\footnote{see \url{https://www.at-bay.com/press_releases/expands-cyber-tech-coverage/}}, while Resilience offers limits up to \$10M for firms with revenue from \$300M to \$10B.\footnote{see \url{https://www.insurancebusinessmag.com/us/news/cyber/resilience-adds-tech-eando-cover-to-cyber-proposition-480136.aspx}}

Given this already existing market, what we describe is not the creation of a new insurance product but rather an evolution and improvement of these policies to account for risk factors identified in this paper. In order to develop proper risk-based pricing, the evolved insurance product should be informed by the previously identified audit, certification, and transparency mechanisms. For example, carriers should be able to collect information about which vendor software products are certified vs. uncertified, which secure coding standards were used (if any), the number of consumer complaints, the vendor's patching cadence, and overall software deployment and maintenance practices.

Ideally, insurance will also play a role in incentivizing risk mitigation behaviors related to software development, management, and deployment practices. Indeed, carriers provide at least four mechanisms for driving risk mitigation practices~\cite{insuring-cat-cyber-risk}. First, carriers will assess the risk posture of the applicant (the software vendor) based on available information (described above) and their own audit practices and, based on that assessment, decide to accept or reject the software vendor's application. Second, as is often discussed for cyber-insurance, carriers can offer price incentives for policyholders to adopt any number of risk mitigation practices that they consider appropriate. Third, carriers distribute ongoing threat intelligence information to firms to help them manage their Internet-facing IT systems, enabling them to preemptively avoid cyber incidents. Fourth, following a claim, carriers provide a panel of firms available to policyholders that can help them manage ex-post losses. We envision carriers can play a similar role with regard to this insurance product. 

On a related note, some believe that with appropriate regulatory support, insurers could actually enforce vendor accountability through their ability to subrogate~\cite{lawfare_Woods24}, i.e., to seek compensation from negligent vendors after paying out the policyholders (customers). Should pro-subrogation regulations be passed (in addition to the liability, transparency, and audit mechanisms discussed above), insurers might become the strong driving force toward a secure software ecosystem as they have been envisioned for the last two decades~\cite{bolot2008cyber}.




\subsection{Regulatory Body}
\label{sec:regbody}

As Section~\ref{sec:state} described, the EU approach relies more on regulation and less on litigation to ensure better outcomes. This is due to a much more powerful regulatory and enforcement arm: the sequence of EU regulations and directives effectively covers some of the components highlighted in Figure~\ref{fig:overview-new}, though not all. By contrast, regulation tends to be much weaker in the US and relies more on litigation and the court system to define appropriate standards of behavior. This is the main reason we have advocated for a system driven mostly by free market forces to align incentives. However, we also believe that some form of light-handed government policy also has an important role to play.

To begin, we believe the software audit office should fall under one of the existing federal agencies, with the most natural fit being the Department of Commerce, potentially as a collaboration between NIST and the Bureau of Industry and Security (BIS). This agency would have the authority to grant certification to the software it audits successfully. The same agency may also host the proposed software clearinghouse. 

Secondly, we advocate for the establishment of an office that collects consumer complaints against software defects and data breaches caused by software defects, just as the FTC's Consumer Sentinel and the FBI's Internet Crime Complaint Center (IC3) collect consumer reports regarding identity theft and fraud reports. These reports would allow consumers to report the nature of the harm they suffered due to faulty software, such as lost work, lost income, physical and mental health issues, etc. Indeed, attempting to measure the actual costs of software defects will greatly help inform our collective understanding of the impact of harms. 

Lastly, a regulatory body could impose penalties of various kinds. For instance, penalties may be assessed for vendors releasing software with known vulnerabilities or vulnerabilities that could be easily detected with common software auditing techniques (e.g., static code analysis, etc). Their detection could be through post-release or post-breach audits commissioned by an insurance company. 


We end this section by briefly commenting on how our framework differs from the current system in the EU.
The EU regulation may be summed up as a combination of (1) mandatory product and organizational audit and certification, and (2) strict product liability (without upper limit). The system we advocate, on the other hand, consists of (i) product/process audits and (ii) the floor-ceiling hybrid approach to liability, which is a weaker version of (2). We also generally believe these should be voluntary, with the possible exception of the floor. An established floor could be made mandatory, meaning no product without a minimal floor audit can enter the market, while the ceiling is voluntary (resulting in liability waiver or capped liability). 
We believe the gap between our liability model with non-mandatory audit and the EU system can be effectively filled by litigation and insurance, as detailed above.

The recently passed EU regulations may also positively affect software vendors opting into voluntary above-minimum product and process audits. As vendors prefer to sell the same software in different regions/markets, they must comply with the mandatory software audit laid out in the EU CRA~\cite{CRA2024} within the next few years. This will have a global effect on software vendors, similar to the GDPR~\cite{bendiek2019externalizing}.

\section{Case Study: the CrowdStrike incident}
\label{sec:casestudies}

\subsection{The Falcon platform}
\label{sec:cs_falcon}
CrowdStrike is a security threat intelligence and software service company whose main product is the Falcon platform, which protects companies against real-time cyber threats. Falcon's architecture comprises a lightweight user agent supported by the CrowdStrike Security Cloud for computational capacity and advanced ML-based analytics powered by large-scale security event data from millions of endpoints worldwide. This enables Falcon to augment traditional signature-based threat detection with real-time anomaly detection for, e.g., malware and network attacks.

\subsection{The incident and its impact}
\label{sec:cs_incident}
On Friday, July 19, 2024, CrowdStrike released a configuration update for its user agent software, installed on Windows computers. An error in the software update caused significant problems for its users: many computers running CrowdStrike services experienced repeated reboots and the notorious Blue Screen of Death (BSOD). The impact of the incident was profound due to the proliferation of Falcon, installed on an estimated 8.5 million Windows machines.

As reported by insurers, the global outage triggered by the defective update from CrowdStrike was estimated to incur financial losses of approximately \$5.4B for U.S. Fortune 500 companies. These estimated losses exclude Microsoft, which experienced extensive system failures during the incident. The aviation, healthcare, and banking sectors were expected to face the most significant financial impacts owing to flight cancellations, disruptions to hospital operations, and widespread payment system failures. According to Parametrix Insurance, insured losses for Fortune 500 companies outside of Microsoft were estimated to range between \$540M and \$1.08B.

This incident is a stark reminder of how a single erroneous software update can cause global operational breakdowns in multiple industries. In fact, our technological ecosystems (particularly those involving software components) are so complex and intertwined that similar disruptions are expected to be the norm rather than the exception.

\subsection{Root cause analysis}
\label{sec:cs_rca}
As thoroughly investigated by CrowdStrike itself~\cite{crowdstrike_rca}, the incident occurred due to multiple issues in CrowdStrike's software testing and deployment processes related to a specific software update.
The Falcon system integrates local (on-sensor) anomaly detection with remote sensor insights via Rapid Response Content, delivered through Channel Files containing various Template Types, which the Content Interpreter processes using regular expressions.

A critical issue arose due to a mismatch in input parameters: a newly introduced inter-process communication (IPC) Template Type required 21 fields, but the integration code supplied only 20. This discrepancy went undetected through multiple validation stages, including sensor release tests and stress tests, as wildcard matching criteria were used for the 21st parameter. Note that this implies a flaw in CrowdStrike's software testing processes.

On July 19, 2024, two new IPC Template Instances were deployed (to provide detection capability for a newly found malware abusing Windows' IPC subsystem), one using a non-wildcard criterion for the 21st parameter. The Content Validator, expecting 21 inputs, failed to detect the mismatch, leading to an out-of-bounds memory read in the Content Interpreter, which the operating system caught, triggering Windows to crash (instead of potentially causing more damage using the garbage data), resulting in the infamous BSOD. Note that Falcon operates in kernel mode (e.g., to provide early boot protection and high performance), hence crashing the entire operating system as opposed to only its own process(es). 

After the crash, the machine either entered a boot loop or booted into recovery mode; either way, manual intervention was needed to restore proper operation. Given the huge number of host machines affected at the same time, the impact was global. Note that the so-called incremental rollout (or canary release\footnote{referring to the use of canaries in coal mines as sentinels for toxic gases}) is an existing software best practice~\cite{tarvo2015canaryadvisor}; by using such a method, the vendor could observe the immediate impact of the newly released update on a smaller set of machines, before gradually pushing the update to more hosts. Such graceful deployment is (would have been) especially beneficial for critical, kernel-level software.

In conclusion, in technical terms, the failure resulted from an input mismatch, an out-of-bounds memory read, a global rollout (three issues could have been prevented by following reasonable software best practices), and Falcon's tight integration with the Windows kernel. Therefore, we posit that this incident occurred because of insufficient quality assurance, manifesting in lax software engineering, software testing, and software deployment processes. 

\subsection{Legal actions}
\label{sec:cs_legal}

There has been significant backlash from customers, shareholders, and regulators\footnote{\url{https://www.securityweek.com/crowdstrike-faces-lawsuits-from-customers-investors/}}. The most prominent attached lawsuit is certainly the one filed by Delta Airlines, which is estimated to have suffered a loss of USD 500 million (not including reputation issues and potential TSA fines) and has had its hands full with more than $176,000$ refund/reimbursement requests. The airline has hired a prominent attorney to pursue potential damages from CrowdStrike and, interestingly, Microsoft. Although Delta is believed to be seeking to recoup all its losses, CrowdStrike moved to dismiss the lawsuit, citing a contractual limit on its own liability and a cap on damages\footnote{\url{https://www.cnbc.com/2024/12/17/crowdstrike-moves-to-dismiss-delta-suit-citing-contract-terms.html}}. Furthermore, CrowdStrike filed a countersuit, blaming Delta's own flawed internal processes for the majority of flight cancellations and refusing their technical assistance during the outage. (Note that other airlines were indeed more effective in handling the outage.) Microsoft made similar comments on the grounds of Delta refusing technical assistance and having an outdated IT infrastructure.

Delta also tried to claim gross negligence or willful misconduct on CrowdStrike's behalf; however, the security vendor argued that converting a breach of contract into tort claims (i.e., product liability) is not feasible under Georgia law (where Delta is based). On the other hand, insurance companies of customers affected by the outage might also go after CrowdStrike, further complicating legal matters. On top of these, CrowdStrike was also sued by its shareholders\footnote{\url{https://www.forbes.com/sites/kateoflahertyuk/2024/08/02/crowdstrike-is-now-being-sued-by-investors/}}, claiming that the technological assurances given by the company were false or misleading, essentially blaming insufficient quality control processes. Note that the company's share price dropped by 44\% in the weeks after the incident. 

Lastly, CrowdStrike's CEO was called to testify in front of the US Congress, and another senior executive apologized multiple times\footnote{\url{https://www.theguardian.com/technology/2024/sep/24/crowdstrike-outage-microsoft-apology}} for letting their customers down, and took ``full responsibility'' for the crashes. Although CrowdStrike conducted an honest forensic root cause analysis\cite{crowdstrike_rca} with public results and vowed to improve their software-related processes, including implementing incremental rollout, evidently, it does not admit legal responsibility, i.e., liability.

\subsection{What if \ldots}
\label{sec:cs_whatif}

One could hypothesize that CrowdStrike's issues stem from insufficient technical expertise; although human error is always a factor, it is far more likely that their simplified software-related procedures were the result of business decisions. They focused mainly on quickly releasing new product features, keeping the customers happy, and their own testing costs low. Simply put, there were no sufficient incentives present to make thorough testing profitable; a textbook incentive issue in software security\footnote{That CrowdStrike is a security vendor makes the situation a bit more striking}. Should our proposed policy framework be adopted, proper incentives for paying attention to (security) testing would emerge: only by passing the product-based floor audit could they market their product; furthermore, successfully passing the voluntary process-based ceiling would result in waived/capped liability, while skipping or failing it would bring with itself full legal responsibility.   

Another important issue is the complexity of software systems in use. In the case study, kernel access created a dependency; while in the incident, an error in the application affected the operating system, this dependency is certainly bidirectional. In the example, our proposed framework would create a dual incentive: CrowdStrike would test its integration into the Windows kernel, while Microsoft could require a minimum quality (i.e., the proverbial floor) from all third-party applications aiming to run in kernel mode. 
As for other production software built on already existing frameworks and libraries, the dependency is unidirectional. Either way, the proposed software clearinghouse is essential for managing security in the software supply chain.

Our policy framework (assuming proper regulations are passed) provides a much clearer legal situation compared to today's US landscape~\cite{lubin_sw_2024}. With the uncertainty gone from legal outcomes, the financial resources of affected stakeholders could actually go into making safer software. Moreover, CrowdStrike's own forensic investigation and pledge to engage independent third parties to conduct a further review of their quality control and software release processes essentially constitute a costly post-incident audit. It is reasonable to assume that an actual pre-incident audit would have been socially optimal, as all impacted stakeholders (CrowdStrike and its shareholders, Microsoft, Delta, and other customers, flight passengers, and other clients of customers, etc.) could have been better off.

Lastly, although CrowdStrike lost 44\% of its value over the 2  weeks after the incident, its share price has almost returned to pre-incident level in the next 2 months\footnote{\url{https://www.investing.com/news/analyst-ratings/crowdstrike-stock-outlook-revised-upward-sector-perform-rating-held-amid-nearterm-risks-93CH-3744295}}. Somewhat surprisingly, the company reported reasonably good 2024 Q3 numbers with an even better outlook for Q4\footnote{\url{https://ir.crowdstrike.com/news-releases/news-release-details/crowdstrike-reports-third-quarter-fiscal-year-2025-financial/}}. They attributed the financial bounce-back to their superior technology, customer commitment packages, and regained reputation. However, even in a market with several competitors, enterprise security software is a sticky product; there is a considerable lock-in effect, especially for larger customers with complex IT infrastructures.  This has significantly contributed to a ``monoculture'' in cybersecurity products; its persistence will only make the next event more costly, more disruptive, and more widespread. Even without knowing the future fortune of the company, such a bounce-back from a major blunder on the back of imperfect competition carries a questionable message. The proposed policy framework could steer the ecosystem in more desirable directions.

\section{Discussions and Limitations}
\label{sec:discussion}

There are other subjects that are often brought up as requiring special attention: the liability of open-source software (OSS) and AI. Below, we explain why these concepts fall naturally under the ecosystem outlined in this paper. We then discuss the limitations and challenges that require further consideration. 

\subsection{Open Source Software}
\label{sec:OSS}

There seems to be a common concern that if/when software liability becomes enforceable, it will have a negative impact on the development of OSS~\cite{lawfare_Dempsey24}. We discuss the potential origins of this concern and whether they are founded.  

It is plausible that the concern is based on the fear that creators of OSS will be held responsible for harm caused by their software, and therefore, any movement in establishing software liability may severely impede innovation and OSS as a major driving force in software development. We do not believe this fear is logically sound. Free stuff (giveaways) never comes with assurances or guarantees, and it is understood {\em user beware} in just about every other domain.  As mentioned in Section~\ref{sec:state}, the EU PLD explicitly excludes free and open-source software outside commercial activity from its regulation, i.e., creators of OSS are free of the liabilities specified by the PLD.  We think this is a very sensible and intuitive way of treating OSS. We believe the liability ultimately lies with the vendor, the one who sells a product for a profit, and one who may have used OSS in its product. By building on OSS, with known or unknown provenance, the vendor effectively assumes all responsibility and liability. Vendors producing and selling critical software or software used in critical infrastructures can decide not to use any OSS, or they can decide to vet any and all OSS used in their product.  As a simple reference, in the US, one cannot give away baby products like car seats or cribs at charity donation centers -- they do not accept these products even for free because they do not want to assume the potential liability that comes with selling or giving away baby products that are tightly regulated for safety. 

It is also possible that the concern is based on the speculation that if a vendor now shoulders all responsibility, perhaps it would be more hesitant to use OSS; an attitude which, if becoming the norm, could potentially discourage OSS development. We don't believe this fear is logical, either. The decision to use or not use OSS in one's production is always driven by economic calculations. While liability can make OSS potentially more costly, that cost increase would similarly apply to one's own software development under a stricter/clearer software liability regime. Furthermore, it is conceivable that an {\em OSS Clearinghouse} could be established by the industry or a government agency, in much the same way as described in Section~\ref{sec:audit}. Such a clearinghouse would essentially assume the responsibility of checking, verifying, and certifying the functionality and safety of OSS. By performing this task, the clearinghouse then assumes (to a certain extent) liabilities associated with the OSS when used downstream.  Perhaps this could be the same agency that does the audit; perhaps there is a difference in the amount of liability assumed by a vendor when it chooses to use an OSS that has not been certified by the clearinghouse. It remains to be seen whether there would be an appetite for such a clearinghouse.  

In short, the type of ecosystem proposed in this paper should not have any real or direct negative impact on OSS development or the innovation that comes with it. 

\subsection{AI (Software) Liability}
\label{sec:AI}


AI risk has often been singled out as a particular risk type needing attention (or special treatment) in both the regulatory circle and among insurance practitioners. This sentiment is clearly tied to the potentially enormous risk and harm that AI products, technologies, and services can inflict. Below, we discuss how AI could be viewed through the lens of the software liability ecosystem outlined in this paper. 

The enormous risk AI presents is most recently exemplified in a case~\cite{FL-teen-lawsuit} involving a 14-year-old in Florida who committed suicide after allegedly being encouraged by an AI chatbot. 
This type of harm notwithstanding, 
we argue that in order to define, recognize, regulate, and mitigate AI risks, one must first do the same for software risks more generally. Without this starting point, we would quickly find ourselves in a situation where we must debate what differentiates AI software from non-AI software and which is governed by what type of regulation. Furthermore, even if there is agreement on the end AI product being distinctly different from other types of software, one must also consider the long chain of software components and mechanisms that go into producing the end product: data acquisition and curation, filtering, basic mathematical algorithms, post-processing, etc. Many of these mechanisms (as implemented in software) are not at all AI-specific -- one finds them in millions of other {\em clearly} non-AI systems.  Where should one draw the line? 

We, therefore, believe that the only tractable way of looking at AI is to first and foremost treat it as software -- there can be no argument that much of modern-day AI is indeed in the form of software (in particular, LLMs and chatbots are all trained, created, and implemented via software mechanisms). It is true that the field is evolving fast, with increasing effort in the form of ``embodied'' AI, which includes hardware (e.g., with sensing, vision, memory, and motion capabilities) as needed in the case of a robot, a quadcopter, and so on (essentially creating a type of cyber-physical system).  However, the ``brains'' of these embodied AI systems (i.e., the data processing, algorithm, and decision making) will continue to exist and operate in the form of software. 

In short, we think it would be helpful and productive for the many stakeholders to first take the high-level view that AI is a particular type of software. We do acknowledge the need in the software liability ecosystem to further carve out a space to address AI liability more specifically. AI systems deserve special attention because they indeed pose new types of risks. For instance, the random and erratic ways LLMs can behave (e.g., in response to prompt engineering) bring additional uncertainty. Another thought-provoking issue is the fact that such systems are only partially ``patchable'' in the conventional sense; new model versions and their dependence and interaction with previous iterations can exhibit deeper and more complex patterns beyond the capabilities of regular software updates. The issues above suggest that AI risks can ultimately differ in profound ways from regular software. 

The set of interconnected EU software/AI regulatory measures (CRA, PLD, AI Act, and the now-revoked draft AILD) aims to achieve AI-specific software liability. In particular, the AI Act~\cite{AIAct2024} classifies AI systems into risk categories and regulates them accordingly. Stakeholders involved (suppliers, importers, deployers, etc.) must adhere to specific requirements, with special emphasis on the high-risk category. This regulation aims to prevent AI-related incidents, while the draft AILD~\cite{AILD} assigned liability in this context. In the U.S., President Trump has rescinded the Executive Order on the Safe, Secure, and Trustworthy Development and Use of Artificial Intelligence (EO 14110~\cite{eo14110}) and signed a new Executive Order on Removing Barriers to American Leadership in Artificial Intelligence (EO 14179~\cite{eo14179}). This move signals a move away from safety and towards minimal barriers to foster innovation and U.S. leadership in AI. 

Interestingly, in the spring of 2025, the EU withdrew the draft AILD, citing the exact ``fostering innovation'' angle. Meanwhile, the California Senate has been discussing SB-813, a bill that proposes a landmark safe harbor for AI developers. This constitutes a (partial) waiver against civil liability claims should the AI developer undergo successful certification conducted by state-approved but independent Multistakeholder Regulatory Organizations (MROs). With international AI legislation seemingly in flux, one might wonder whether an alternative approach, a type of Hippocratic oath for AI developers~\cite{sharma2024ai}, could be successful in the face of minimal regulations and ineffective litigation against big tech.

\subsection{Limitations}
\label{sec:limitations}

While we believe the framework we have outlined is sound and can effectively help realign misaligned incentives, we also acknowledge that ``the devil is in the details'': much remains to be worked out for each of the components in the ecosystem.  In particular: 
\begin{itemize}
    \item Converging on a floor-safe harbor legal regime will require more specific definitions of the floor and the safe harbor; scholars will need to agree on what constitutes a minimum requirement and what constitutes reasonable best effort. 
    \item How to perform a software audit is another major open technical area: is it possible to establish a set of audit standards, how to ensure this does not become yet another checklist, when a certain software has been certified to enter the clearinghouse, is it supposed to be safe for all use cases?  
    \item How to reconcile the discrete (one-time) nature of audit and the far more fluid nature of software development: can a vendor submit one version for audit but then release onto the market a different version (e.g., by turning off privacy-enhancing features)? 
    \item How to balance the accuracy and efficacy of an audit with the privacy/proprietary needs of the vendor?
    \end{itemize}
    
    In this sense, we view this paper as a call to arms: this effort will require the engagement of a large research community, and the time to act is now. 



\section{Conclusion} 
\label{sec:conclusion}
Making software products more reliable and more secure is critical to our increasingly digitized society. As we have witnessed recently, even minor (and non-malicious) software failures can paralyze entire industries, such as airline and stock trading, wreaking havoc on a global scale with enormous business disruptions and economic impact. More serious vulnerabilities enable intentional, malicious attacks that can create equally, if not more devastating, consequences. 



We envision an ecosystem where the concept of software liability may be used effectively and responsibly to simultaneously (1) incentivize better and more secure software development and (2) reallocate some of the costs (e.g., that result from data breaches attributable to bad/insecure software) currently borne almost entirely by consumers.  

We highlighted how we think such an ecosystem can function and how our proposal affects incentives among the (old and new) stakeholders: software developer/vendor, consumer, software auditor, litigator, insurance underwriter,
and how they each have the incentive to be part of this ecosystem.  

As a community of researchers and practitioners, we must not stop seeking the right balance and tradeoff: the single-minded pursuit of innovation may or may not justify the mounting collateral damage -- the potential harm caused by software products has become crystal clear in this age of entangled software supply chain and AI.
We believe this is a critical time to reassess the question of whether and how we can demand better quality software as a society. 

In a very recent and positive development, the outgoing U.S. administration passed Executive Order 14144~\cite{eo14144}, which builds on EO 14028~\cite{whitehouse_2021} to strengthen cybersecurity and promote (software) innovation at the same time. Among others, the new EO sets out mandatory software security requirements for suppliers of the federal government: 
software providers must submit machine-readable attestations of secure development practices, along with validation artifacts, to the Cybersecurity and Infrastructure Security Agency (CISA). If such stringent requirements can spill over to the non-government software market, they have the potential to catalyze the proposed incentive realignment. Note that this EO has not been rescinded, seemingly affirming that software security is a bipartisan issue.

\section*{Acknowledgments}
Gergely Biczók was supported by the Fulbright Visiting Scholar program and 
Project no. 138903 of the Ministry of Innovation and Technology, Hungary, from the NRDI
Fund, financed under the FK\_21 funding scheme. 
Mingyan Liu was supported by the US National Science Foundation (NSF) under grant IIS-2112471. 

\bibliographystyle{unsrt}
\bibliography{audit_refs}

\appendix
\section{Appendix: Detailed Literature Review}
\label{sec:literature}

In this section, we provide a more comprehensive review of software liability and the state of the European and American legal landscape around the matter for the interested reader. 

\subsection{Software liability}
\label{sec:sw_liability}
Ex-post liability is built on waiting for an adverse event to occur, then assigning liability to the injurer~\cite{cybok-security-economics-kg}. The legal interpretation of liability already appeared in Roman Law, but the modern \emph{framework of product liability} first emerged in the early twentieth century~\cite{macpherson_1916} in relation to a faulty automobile wheel causing injuries to the driver. The ruling cited the reasonable basis for the automobile manufacturer to know product risks and to make an extra effort to ensure the safety of anyone coming into contact with the car, even if the wheel was produced by another company. The next milestone case in 1932~\cite{donoghue_1932} established \emph{duty of care} with respect to a person who consumed spoiled ginger beer produced by a bottler and contracted severe gastroenteritis. The bottler had to pay restitution owing to his negligence. Finally, \emph{standard of care} was put forward in 1947 ~\cite{us_1947} in a complicated multi-party liability case among the US government, a railroad company, and two boating companies, where the mooring lines of a barge carrying flour belonging to the government were disconnected by another barge, causing damage to other ships and eventually sinking. Liability was shared among the companies; specifically, the boating company operating the barge that cut the lines failed to look after the safety and security of the other barges affected. Interestingly, the judge based his ruling on \emph{inadequate risk management}: the burden of precautions taken by the company was smaller than the likelihood, and the impact of the incident multiplied together. 

Although there were several liability cases against software companies in the last decades, companies often escaped fines using nuances of contract law and limitations of liability and disclaimers in end-user license agreements (EULA)~\cite{lubin_sw_2024}. In fact, the flexible interpretation of acceptable risk taken by the user of a software product made this a gray area. This has changed dramatically towards the end of the last decade with the emergence of the European General Data Protection Regulation (GDPR) and its penalty system for organizations mismanaging the personal data of their users. The GDPR has defined both the duty and standard of care, data protection agencies have collected billions of Euros in fines, and the cost-benefit assessment of cybersecurity-related investments has changed for the better, at least in terms of the safekeeping of personal data. Concurrently, governments worldwide have become increasingly worried about the changing landscape of cyberattacks, which has shown a shift from accessing/exposing user data towards damaging and controlling critical infrastructure integrated across governmental institutions and private organizations. The 2017 NotPetya ransomware~\cite{notpetya_2018} and the 2021 SolarWinds attacks~\cite{peisert2021perspectives} stand out as having far-reaching implications for national security, letting malicious attackers breach and potentially take over control of systems of critical importance (governmental organizations, financial institutes, logistics companies, etc.) in the Ukraine, in the United States, and across the globe. These incidents made governments prioritize the cyber-hardening of their national infrastructure and, specifically, focus on software supply chain security. Accordingly, both the United States (via the 2021 Executive Order on Improving the Nation’s Cybersecurity~\cite{whitehouse_2021}) and the European Union (via the 2022 Cyber Resilience Act~\cite{eucyber_2022}) acted swiftly and put forward regulations mandating secure software development, standardized product cybersecurity compliance, and a secure software supply chain~\cite{sonatype_2023}, and also reducing the information asymmetry pertaining to non-expert end-users of software products. Software liability also factors in the end-user's technical (lacking expertise to assess product security) and legal inability (EULA prohibiting the inspection of software code) to evaluate software products they use.
In fact, the EU has enacted a new Product Liability Directive \cite{PLD2024} that came into force on 9th December 2024, with the intent to explicitly designate software as a ``product''; this, of course, comes with strict liability clauses. The recent CrowdStrike incident has shown the world that this just might be an effective approach for managing such scenarios (see Section~\ref{sec:casestudies} for a detailed case study).

Another intriguing aspect of software liability comes from cyber-insurance. The devastating cybersecurity incidents and the ongoing multi-modal armed conflicts affecting critical infrastructure raised alarms in the insurance industry regarding the insurability of cyber risks~\cite{wef_2023}, potentially inhibiting the market that experienced real growth in recent years after decades of underdevelopment. In fact, insurers, reinsurers, academics, and cybersecurity experts are trying to establish baseline scenarios in the industrial control system domain~\cite{dagstuhl_2022}. Indeed, if the cyber risk is systemic and catastrophic losses are expected in the critical infrastructure sector, the insurance industry will lose interest in underwriting new policies. In that case, governments would shoulder the responsibility to be insurers of last resort, similar to instances of frequent natural disasters~\cite{cummins2006should}. Supposing this happens, the increased scrutiny for software product security laid down in recent regulations is even more sensible~\cite{lawfare_herberstein}. Even so, a potential insurer role would also be a financial and administrative burden for the government, especially taking into account tangled software supply chains. On the other hand, supply chain security solutions and best practices are available; if a company decides to invest money and effort, a much-improved level of security is within reach. Recognizing this, the current United States National Cybersecurity Strategy~\cite{house2023national}, released in April 2023, proposed a liability waiver mechanism designed to reward software companies willing to undergo and pass a government-mandated product security audit. (Such audits have been studied extensively; we refer the reader to~\cite{huang_incentivizing_2024} for an overview.) Such a waiver mechanism could just be the missing financial incentive for software companies to change their usual ways regarding secure products.

\subsection{European Regulation}
\label{sec:eu}

\begin{figure}[tb]
\centering
\includegraphics[width=0.99\textwidth]{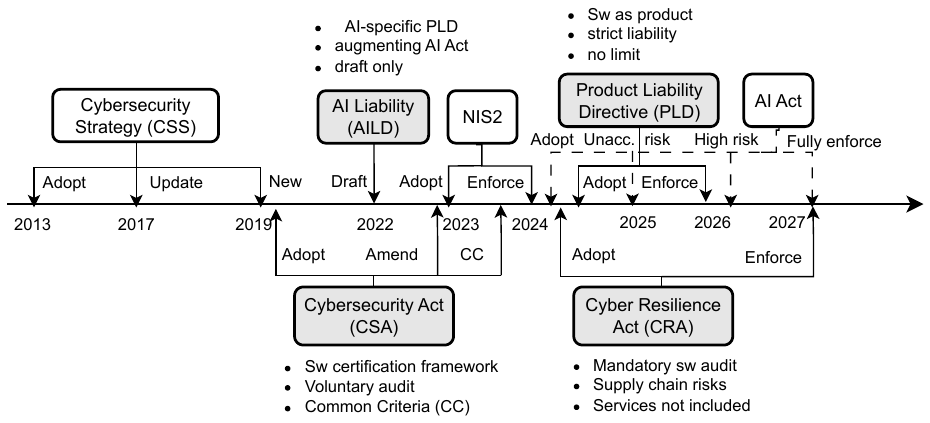}
\caption{Timeline of key EU regulations and highlights of their contents (regulations directly affecting software liability and audit in gray).}
\label{fig:EU_regulation_timeline}
\end{figure}

The European Union provides an important comparison in regard to software regulation and management. For example, the EU Cybersecurity Strategy~\cite{EUstrategy2020} includes many acts and directives aimed at governing cybersecurity-related activities and together creates a holistic set of initiatives to incentivize better software development. Some of these regulations are already in effect; some have been put into force, but they provide a grace period for affected stakeholders to comply, and some are still in the draft phase. Below, we present the regulations that directly or indirectly impact (software) liability. A timeline of key regulations and their main components is presented in Figure~\ref{fig:EU_regulation_timeline}. 

\noindent \textbf{Cybersecurity Act, 2019/2023. }The first step in the regulatory initiative was the 2019 Cybersecurity Act (CSA)~\cite{CSA2019}. Most importantly, the CSA established the EU Cybersecurity Certification framework\footnote{\url{https://digital-strategy.ec.europa.eu/en/policies/cybersecurity-certification-framework}} (EUCC) for ICT products, services, and processes. The EUCC provides risk-based EU-wide certification schemes for ICT products/services as a comprehensive set of rules, technical requirements, standards, and procedures. Each certification scheme, the first one implemented being the widely used Common Criteria~\cite{EUcommon2024}, defines the categories of products and services covered, the cybersecurity requirements, the types of evaluation required (self-assessment vs. third-party audit), and intended assurance levels. The assurance level (basic, substantial, or high) informs users of the cybersecurity risks associated with a product, reflecting the probability and impact of potential incidents. A high assurance level indicates the product has passed strict security tests by an authorized third-party auditor, and the resulting certificate will be recognized across all EU Member States, facilitating cross-border trade and transparency for purchasers. National authorities 
oversee the certification process in each state, while ENISA is tasked with the EU-wide coordination. Note that \emph{certification is voluntary} in the scope of CSA. The CSA received a targeted amendment proposal on managed security services in 2023~\cite{CSAamend2023}, expanding the certification coverage of EUCC to incident response, penetration testing, \emph{security audits}, and consultancy. 

Managed security service providers, critical due to their integration with customer operations, are also classified under the \emph{high criticality} sector in the 2022 NIS2 Directive~\cite{NIS2}, transposed into national law by October 17, 2024. NIS2 promotes security across key sectors like energy, transport, banking, and healthcare. It mandates supply chain security measures and \emph{compulsory security audits} for organizations in these sectors.

\noindent \textbf{Cyber Resilience Act, 2024. }
The Cyber Resilience Act (CRA)~\cite{CRA2024}, effective from December 10, 2024, mandates cybersecurity requirements for manufacturers and retailers of digital products (PDEs). It ensures built-in security throughout the product lifecycle (\emph{duty of care}), addressing weak cybersecurity and helping consumers identify secure products through harmonized rules and obligations.

PDEs are categorized into three risk-based groups. Low-risk products (e.g., smart speakers) require basic compliance via self-assessment. “Critical class I” products (e.g., password managers) must meet stricter standards through recognized ISO/IEC/ETSI certifications or independent audits. The highest-risk PDEs require mandatory third-party audits. Note that CRA compliance is mandatory, but EUCC-certified PDEs will automatically comply.

The CRA emphasizes supply chain security for third-party components in PDEs. Component manufacturers selling in the EU must comply, and PDE manufacturers must ensure secure sourcing and report vulnerabilities. Compliance falls on the importer for non-EU components if the manufacturer doesn't sell directly. The CRA mandates Software Bill of Materials (SBOM) as an essential artifact for vulnerability tracking, with the German cybersecurity authority already providing implementation guidelines~\cite{BSIsbom2024}. While digital services like SaaS are excluded, the NIS2 Directive neatly complements CRA by setting cybersecurity and reporting requirements for critical providers of those services.

PDEs under strict sectoral regulations (e.g., cars, medical devices) are exempt from the CRA. Open Source Software (OSS) is also excluded if not monetized, as it doesn’t qualify as a commercial activity. Further clarification confirms that funding and development circumstances don't affect OSS's commercial status\footnote{\url{https://openforumeurope.org/eu-cyber-resilience-act-takes-a-leap-forward/}}.

\noindent \textbf{New Product Liability Directive, 2024. }
The new Product Liability Directive (PLD)~\cite{PLD2024} has replaced the 1985 directive, which introduced a no-fault strict liability regime in the EU. The update modernizes product liability rules for the digital age. The new PLD allows individuals, including consumers, to seek compensation from manufacturers on a strict liability basis for defective products and, in some cases, their components in the EU market. Its goal is to simplify the claims process for damages caused by product defects.

The PLD expands liability to nearly all supply chain operators and ensures consumer protection regardless of a product’s origin (whether EU or non-EU). Even online marketplaces may be liable if they act like sellers, but they can avoid it by providing details of the manufacturer's EU representative. Consumers will gain access to claim-related information while maintaining confidentiality. They can seek compensation in complex cases, including safety/security regulation breaches (e.g., CRA or AI Act). The removal of arbitrary thresholds allows full compensation for damages.

The New PLD defines ``product'' broadly, covering physical goods, raw materials, and standalone software, including AI software. It addresses new technologies such as cyber vulnerabilities, essential digital services, and software updates. Note that free and open-source software outside commercial activity is excluded. “Component” includes any item (e.g., a software library), raw material, or service (e.g., when software initiates remote calls to a SaaS instance) integrated into or connected with a product, though standalone services are generally not covered.

\noindent \textbf{AI liability. }
Artificial intelligence has rapidly advanced over the past decade, becoming a key digital technology. In response, the EU introduced the AI Act~\cite{AIAct2024}, coming into force in 2024, and drafted the AI Liability Directive (AILD)~\cite{AILD}. The EU recognized that even under the new PLD, victims must prove fault, damage, and causation, especially a difficult task given AI’s complexity and autonomy. Ensuring victims can access compensation is crucial, and AI’s opacity should not hinder justice.

The AI Act and the AILD are complementary components of a unified approach: they operate at different stages but strengthen each other. The AI Act focuses on minimizing risks and preventing damage through safety-oriented regulations. However, since it is impossible to eliminate all risks, the AILD ensures that if damage does occur, compensation remains effective and feasible. Essentially, while the AI Act aims to avert damage, the AILD provides a safety net for compensation when damage does happen.

The AILD aligns with the AI Act by adopting its definitions and risk classifications. It builds on the AI Act's transparency requirements, ensuring actionable provisions for liability, such as the right to information disclosure. The Directive applies to AI-caused damage, regardless of risk classification.

The new PLD is limited to claims brought by private individuals. In contrast, applying also to legal persons, the draft AILD reforms national fault-based liability systems, covering claims against parties responsible for AI system faults causing damage, including discrimination or privacy violations. Note that Directives introduce mechanisms like evidence disclosure and rebuttable presumptions to ease the burden of proof and maintain coherence across compensation processes.

The AILD has been in a draft phase, with the EU promising the finalization of proposed amendments\footnote{\url{https://datamatters.sidley.com/2024/03/21/eu-formally-adopts-worlds-first-ai-law/}}, while the AI Act and the new PLD are already in force. In a somewhat unexpected turn of events, the EU announced the withdrawal of the AILD in February 2025 together with releasing their 2025 Work Programme. The withdrawal elicited mixed responses with some expressing significant concern while others welcomed legal simplification and targeted deregulation\footnote{\url{https://www.euractiv.com/section/tech/news/commission-plans-to-withdraw-ai-liability-directive-draw-mixed-reactions/}}, enabling AI-based European innovation.

As AI systems involve complex software (and hardware) with AI-specific dependencies, vendors and stakeholders are advised to take software liability seriously.

\noindent \textbf{Summary. }The EU devised an elaborate and strict legal framework, ``throwing a hand grenade''~\cite{lawfare_uren} on explicit software liability, (partially) including AI-enabled software. The situation is quite different in the US.

\subsection{US Regulation}
\label{sec:us}

As mentioned above, the current U.S. National Cybersecurity Strategy~\cite{house2023national} put forward the concept of software liability together with an audit-based liability waiver mechanism. While the significance of this step has not gone unnoticed in both the legal and tech communities, its codification and application are not without challenges. 
 
\noindent \textbf{Issues with US tort law. }
Software is not classified as a product or product component under current US law~\cite{lubin_sw_2024}. This has to do with the perceived intangible nature of software vs. the definition of product as ``tangible personal property distributed commercially for use or consumption.'' Court rulings largely align with this view by treating software as a service rather than a product, forcing plaintiffs to rely on legal arguments of negligence liability when bringing private actions against software providers. Given the ubiquity of software, the potential for harm is boundless, and duties and standards of care are hard to establish since the same software code could be utilized in different applications and use cases. 

It is safe to say that negligence alone is not an effective framework for software in itself; product liability might be a better fit. In strict liability, attention is on the product, and there is a long history of solving complex disputes in other domains, such as health and automotive. The existence of ``component liability'' and ``supply chain risk'' (dating back to the case of the automobile wheel~\cite{macpherson_1916}) is a match made in heaven for software. Another key tool is expert testimonies, which are needed to evaluate technical cybersecurity debates.

Further complicating the matter are two major factors. First, the fragmented strict liability schemes in the 50 states either do not define ``product'' clearly or define it via traits of ``tangibility'' followed by a requirement on commercial purpose. Second, most of the existing case law holds that software is not a product. Treating software vendors as content providers, courts usually designate software as a service (thus not falling under strict liability), as evidenced by decisions in cases against Airbnb, Uber, Facebook, Snapchat, and Amazon~\cite{lubin_sw_2024}. Majority notwithstanding, in a couple of recent cases, software (e.g., the Lyft app) or components of software (e.g., parental control functionality in social media apps) were deemed as products/component products based on explanations that the definition of ``product'' should accommodate technology developments and that intangible personal property might even qualify as such. These decisions might signal a judicial change.

\vspace{1mm}
\noindent \textbf{Liability or Transparency? }
While the U.S. legal landscape is unclear on the matter, there is no consensus about whether a liability regime is the ``right path'' forward~\cite{darkreading_Williams23}. Scholars argue that although market failures necessitate some form of intervention, a transparency-based approach might have the same positive effect while still avoiding increased costs, slower innovation, stifled competition, and present scalability issues. Mandating full disclosure of security practices removes information asymmetry and could enable consumers to assess risks and create a market-driven demand for more secure software. In theory, this would make secure development efforts financially viable (or even desired). Indeed, Lipner argues that Executive Order 14028~\cite{whitehouse_2021} has already laid out a transparency framework (based on self-attestation to the adoption of secure development practices)~\cite{lawfare_Lipner24}. Some think that attestation to a government-mandated requirement as part of procurement could incentivize vendors to improve their software development and deployment practices.

There seems to be a consensus among scholars that attestation will indeed play an important role~\cite{lawfare_Dempsey24b}. In addition to self-attestation, a cybersecurity ``safe harbor'', based on the certification of organizations (with a similar approach already codified in the EU NIS2 Directive~\cite{NIS2}, has received considerable attention~\cite{lawfare_Tschider24}. Proponents argue that a certification safe harbor has the ability (over time) to create incentives to improve collective security practices, including supply chain security, by involving certified upstream technology providers. 

Dempsey suggests a hybrid approach, combining a product-based ``floor'' for liability and a process-based ``ceiling'' for safe harbor~\cite{lawfare_Dempsey24}. His proposed mechanism for liability is three-fold. First, a rule-based approach would define the minimum standard of care for software, focusing on product features and behaviors to be included or avoided (e.g., any weaknesses tracked in the MITRE Common Weakness Enumeration (CWE) database\footnote{\url{https://cwe.mitre.org/}}). Second, as software can be complex and dynamic, a liability regime and analysis are needed to assess non-trivial design and implementation flaws; essentially, a software audit. Third, to provide a safe harbor to developers who followed secure coding recommendations, the proposal suggests a liability waiver or limitation, which makes the outcome predictable for vendors. Clearly defining the conditions under which a vendor qualifies for safe harbor provides legal certainty, reducing the need for costly and prolonged expert debates over what qualifies as reasonable (non-negligent) conduct~\cite{lawfare_Bambauer24}. 

\noindent \textbf{Regulating Open Source Software. }
The 2021 Executive Order on Improving the Nation’s Cybersecurity~\cite{whitehouse_2021} catalyzed an initiative in open-source security, leading to the Open Source Software Security Mobilization Plan~\cite{oss_security_plan}. This plan outlined ten priority areas, including training, digital signatures, and Software Bill of Materials (SBOM) requirements. A significant legislative step followed with the introduction of the bipartisan Securing Open Source Software Act~\cite{oss_security_act}, which defines the Cybersecurity and Infrastructure Security Agency’s (CISA) role in enhancing open-source security. Arriving in the footsteps of the Log4j debacle~\cite{enck2022top}, the proposal marks a pivotal moment in U.S. policy, recognizing open source as essential to national security and emphasizing the government's supporting role in its long-term resilience.

The Act aims to assign key responsibilities to the CISA director, emphasizing collaboration with government entities, the private sector, and open source organizations to enhance long-term open source security. It mandates the development of a risk assessment framework for critical open source components, covering component identification, secure development processes, and SBOM creation to track vulnerabilities. The act also establishes guidance for government CIOs based on open-source best practices, aligning with the growing trend of Open Source Program Offices (OSPOs) to manage risks and contributions. A pilot OSPO initiative will help define government policies for secure open-source engagement.

As evidenced by continuously emerging critical vulnerabilities in popular open source software, such as CVE-2024-3094\footnote{\url{https://nvd.nist.gov/vuln/detail/CVE-2024-3094}} in the Linux compression utility XZ Utils, open source software security, especially in the supply chain context, is crucial. However, we have two additional observations. First, bugs and vulnerabilities in proprietary software could be just as critical. Second, in terms of liability, open source should be excluded (similarly to the EU CRA~\cite{CRA2024}), not to thwart innovation; however, commercial products built on top of open source libraries should be under scrutiny. Note that the Act itself has not yet been enacted.


\noindent \textbf{Regulation vs. insurance. }
Woods examines the role of insurance in the proposed software liability regime, highlighting its potential to both undermine and support policy goals~\cite{lawfare_Woods24}. On one hand, insurance could weaken incentives for security improvements if software vendors simply transfer liability costs to insurers without changing their development practices, a classic case of moral hazard. 

On the other hand, insurance can positively strengthen software security and ensure victim compensation. Cyber liability insurance already helps businesses cover costs associated with security breaches, providing a financial safety net for affected parties. Furthermore, liability insurance can offer predictability and stability for vendors by managing litigation risks, allowing them to better navigate potential legal challenges. In some cases, insurers may even encourage stronger security practices by requiring vendors to adopt safer development processes to qualify for coverage or lower premiums.

Woods hypothesizes that, under the right regulatory regime, insurers can act as key enforcers of accountability. Currently, for example, if a hospital suffers a ransomware attack due to flaws in its online authentication system, its cyber insurance would cover the costs, but the software vendor responsible for the security failure would likely escape liability. This raises concerns about justice and accountability, as the legal system typically assigns costs to the negligent party. To address this, insurers should have the right to subrogate, meaning they can seek compensation from the negligent vendor after covering the victim’s losses. Subrogation serves two key principles: holding wrongdoers accountable by ensuring they bear the financial consequences of their negligence and upholding the indemnity principle, which prevents victims from profiting through double recovery from both an insurer and a liable party. By integrating subrogation into the liability regime, insurers can reinforce both fairness and security in software accountability.

Woods proposes that lawmakers strengthen the insurers’ ability to subrogate against software vendors by 
\begin{itemize}
    \item Invalidating waivers of subrogation: these contract clauses block insurers from recovering costs, reducing vendor accountability. Removing them enhances liability enforcement.
    \item Ignoring insurance payouts in liability rulings; the collateral source rule ensures vendors remain fully liable, preserving their incentive to improve security.
    \item Clarifying award distribution; prioritizing insurer recovery in cyber cases deters insecure software and aligns with public policy goals (of a more secure software ecosystem).
\end{itemize}

\noindent \textbf{The recent Executive Order 14144. }
Before leaving office, President Biden issued EO 14144 on Strengthening and Promoting Innovation in the Nation’s Cybersecurity~\cite{eo14144}.  The EO signals a shift from voluntary cooperation toward mandatory security mandates, placing greater responsibility on software and cloud service providers rather than end users. It addresses critical vulnerabilities in the digital infrastructure, underscored by recent high-profile security breaches. It leverages procurement authority, economic sanctions, mandatory agency practices, and non-binding standards to strengthen cybersecurity. By imposing new federal contract requirements, the EO is expected to influence the private sector by promoting secure products and increasing market transparency.

Notably, as related to the software lifecycle, first, the EO recognizes the need for security not only in the development of software but also in the delivery of software and the application of patches. Second, it also requires federal government vendors to attest to the security of their development practices, providing more (self-reported) transparency. Third, these attestations will be validated by the Cybersecurity and Infrastructure Security Agency (CISA); these validations will be publicly available; therefore, non-government customers will also benefit from the added transparency in their purchasing decisions. What's more, vendors whose attestations fail validation may be referred to the Department of Justice; this brings similarities with the corresponding EU legislations, such as NIS2. Given its largely bipartisan nature, the EO has a good chance of staying in effect.

\end{document}